\documentclass[aps,prb,showpacs]{revtex4}
\usepackage{epsfig}
\newcounter{tr}
\newcommand{\bqe}{\begin{eqnarray}}
\newcommand{\eqe}{\end{eqnarray}}

\newcommand{\sfr}{^{\frac{1}{2}}}
\newcommand{\vl}{\vec{l}}

\newcommand{\vQp}{\vec{Q}_{\|}}
\newcommand{\la}{\langle}
\newcommand{\ra}{\rangle}

\begin{document}
\title{ Ferromagnetic Resonance Linewidths in
Ultrathin Structures; Theoretical Studies of Spin
Pumping}
\author{A. T. Costa$^1$, Roberto Bechara Muniz$^2$ and D.L. Mills$^3$}
\affiliation{$^1$ Departamento de Ci\^encias Exatas\\
Universidade Federal de Lavras\\
37200-000 Lavras, M. G. Brazil}
\affiliation{$^2$ Instituto de F\'{\i}sica\\
Universidade Federal de Fluminense\\
24210-340 Niter\'oi, R. J. Brazil}
\affiliation{$^3$ Department of Physics and Astronomy,
University of California, Irvine, CA 92697, U. S. A.}
\date{\today}
\begin{abstract}
We present theoretical studies of the spin
pumping contribution to the ferromagnetic
resonance linewidth for various ultrathin film
ferromagnetic structures. We consider the
isolated film on a substrate, with Fe on Au(100)
and Fe on W(110) as examples. We explore as well
the linewidth from this mechanism for the optical
and acoustical collective modes of FM/Cu$_{\rm
N}$/FM/Cu(100) structures. The calculations
employ a realistic electronic structure, with
self consistent ground states generated from the
empirical tight binding method, with nine bands
for each material in the structure. The spin
excitations are generated through use of the
random phase approximation applied to the system,
including the semi infinite substrate on which
the structure is grown. We calculate the
frequency response of the system directly by
examining the spectral density associated with
collective modes whose wave vector parallel to
the surface is zero. Linewidths with origin in
leakage of spin angular momentum from the
adsorbed structure to the semi infinite substrate
may be extracted from these results. We discuss a
number of issues, including the relationship
between the interfilm coupling calculated
adiabatically for trilayers, and that extracted
from the (dynamical) spin wave spectrum.  We
obtain excellent agreement with experimental
data, within the framework of calculations with
no adjustable parameters.
\end{abstract}
\maketitle

\section{Introductory Remarks}

The damping of spin motions in nanoscale
ferromagnetic structures has been a topic
explored actively in recent years. The interest
focuses on the 3d transition metal ferromagnets
and their alloys, since one may now synthesize
diverse ultrasmall magnetic structures of very
high quality from them. It is the case as well
devices are fabricated from these materials, in
which ferromagnetism is realized even well above
room temperature. It is very clear that one
encounters damping mechanisms not present in the
bulk crystalline form of the constituents. That
this is so has been evident for many years
now\cite{1}. The primary question then centers on
the origin of the new damping mechanisms
operational in nanomagnetic structures.

Of course, the electronic structure of such
entities may differ substantially from that in
the bulk crystalline matter, by virtue of
distortions of the lattice associated with the
mismatch in lattice constant between the film and
the substrate. In addition, a large fraction of
the moment bearing ions sit at interfaces or on
surfaces, and this will influence the electronic
structure as well. Hence, if the Gilbert damping
constant $G$ is used as a measure of the damping
found for long wavelength spin motions, one may
expect intrinsic differences between the values
of $G$ appropriate to nanoscale structures, and
that appropriate to bulk materials.

It is now very clear from diverse experimental
studies that the damping mechanisms operative in
ultrathin ferromagnets do not have their origin
only in differences between the electronic
structure of bulk and nanoscale matter. Evidently
distinct new mechanisms are present. For
instance, an earlier analysis of data on
ferromagnetic resonance (FMR) linewidths show a
dependence on growth conditions whose influence
on electronic structure is surely rather
indirect\cite{1}. A few years ago, it was
argued\cite{2} that a mechanism referred to as
two magnon damping can be activated by defects on
the surface or at interfaces. The density and
character of such defects is clearly influenced
by the manner in which the film is grown. The
theory of two magnon damping makes explicit
predictions regarding the frequency variation of
the linewidth and its dependence on the
orientation of the magnetization\cite{2,3}. Also,
Rezende and his colleagues have demonstrated that
it accounts quantitatively for the strong wave
vector dependence found upon comparing linewidths
measured in FMR, and those measured in Brillouin
light scattering (BLS)\cite{4}. We refer the
reader to a review article which describes the
theory of two magnon damping in ultrathin
ferromagnetic films, and the experimental
evidence for its presence\cite{5}.

Two magnon damping is an extrinsic mechanism
activated by defects on or within an ultrathin
film. In the recent literature, experimental
evidence has been presented for the presence of a
mechanism operative in ultrathin ferromagnetic
film structures which is intrinsic in
character\cite{6}. It should be remarked that
Berger\cite{7} and Slonczewski\cite{8}  predicted
this mechanism should be present in ultrathin
films in advance of the experiments. Suppose we
consider an ultrathin film of a ferromagnetic
metal placed on a metallic substrate. The
magnetic moments in an ultrathin ferromagnetic
film are embedded in a sea of conducting
electrons which, of course also have spins and
magnetic moments. As the magnetic moments are
excited in an FMR or BLS experiment, they precess
coherently. Angular momentum is transferred to
the conduction electrons, and this is transported
across the interface between the film and the
substrate in the form of a spin current. Thus,
spin angular momentum is lost from the
ferromagnetic film, and the spin motions are
damped as a consequence. This source of damping
is referred to as the spin pumping mechanism. As
the ferromagnetic film is made progressively
thicker, the spin pumping contribution to the
linewidth decreases roughly inversely with the
film thickness, according to theory. This
behavior is found in the experimental data.

By now, numerous theoretical descriptions have
been given for the spin pumping contribution to
the linewidth. In the early papers cited in the
previous paragraphs\cite{7,8} a simple
phenomenological picture of the ferromagnetic
metal and the surrounding materials is employed.
The magnetic moments are described as localized
moments, with $\vec{s}(\vec{l}$) the spin angular
momentum associated with lattice site $\vec{l}$.
Such local moments are assumed to be coupled to a
conduction electron bath, modeled as free
electrons in quantum wells.  These interact with
the local moments via the classical sd exchange
interaction $-J\vec{S}(\vec{l})\cdot
\vec{\sigma}$, with $\vec{\sigma}$ the conduction
electron spin.  Within such a framework, general
features of the spin pumping mechanism may be
elucidated, but quantitative predictions for
specific materials are not contained in such
simple descriptions.

Within such a framework, Simanek and
Heinrich\cite{9} introduced an elegant picture of
the origin of the spin pumping contribution to
the linewidth. A ferromagnetic film with static
magnetization placed in contact with a non
magnetic metal induces RKKY spin oscillations in
the non magnetic material. If the magnetization
of the ferromagnet precesses at a finite
frequency, the RKKY oscillations do not quite
follow the magnetization of the substrate. In
essence, the RKKY coupling is frequency
dependent. If the frequency dependent RKKY
interaction is expanded in powers of the
frequency, the lowest order linear term in the
effective equation of motion of the magnetization
provides a contribution to the effective damping
constant felt by the ferromagnetic spins. In Ref.
[9], only a single layer of localized spins was
considered, embedded within an electron gas. One
of the present authors has elaborated on this
basic scheme by considering $N$ layers of local
moments to simulate a multilayer film, with
conduction electrons described by a quantum well
picture\cite{10}. A prediction which emerges from
this model is that the spin pumping contribution
to the linewidth does not fall off simply like
$1/D$, with $D$ the thickness of the
ferromagnetic film. Quantum oscillations are
superimposed on this
$1/D$ falloff.  Further discussion of this
approach has been presented by Simanek\cite{11}.

A rather different approach has been taken by
Tserkovnyak and his colleagues\cite{12}. These
authors use a one electron picture, in which the
electron wave functions extend over the entire
structure considered, the ferromagnetic film and
the non magnetic metals which are in contact with
it. Spin precession within the ferromagnetic film
is introduced by virtue of a postulated form for
a time dependent density matrix. The flow of spin
angular momentum out of the ferromagnetic film in
the presence of the spin precession is described
in terms of the transmission and reflection
coefficients associated with appropriate
scattering states. Recently a study by this group
has appeared where density functional electronic
structure calculations are employed to calculate
spin pumping contributions to the linewidth for
specific material combinations\cite{13}. The
agreement with experimental data which follows
from this analysis is very good, though we note
that the theory explores an ultrathin
ferromagnetic film bounded on both sides by a
semi infinite non magnetic metal, a structure
quite different than employed in the actual
experiments. The authors conclude that the
oscillations found within the quantum well model
of Ref. [10] are in fact very modest in amplitude
when a proper electronic structure is employed.
We remark that in the series of calculations
which motivate the present paper, we find similar
results.

In this paper, we address the calculation of spin
pumping linewidths for various ultrathin
film/substrate combinations by a method quite
different than found in the studies cited above.
We calculate directly the full frequency spectrum
of the response of diverse structures to applied
microwave fields whose wave vector parallel to
the surfaces of the structures is zero, within
the framework of a method which provides a
realistic description of the electronic structure
of the sample of interest. We extract linewidths
by fitting the lines in the absorption spectra
with a Lorentzian, very much as done in
experimental analyses of actual data. As remarked
above, we note that the formalism employed by the
authors of Ref. [12] and Ref. [13] requires the
ferromagnetic film to be bounded on both sides by
non magnetic metals of semi infinite extent,
since the reflection and transmission
coefficients required are found from scattering
states associated with electrons reflected from
and transmitted through the ultrathin
ferromagnetic film. Our method allows us to
address the more realistic case of an ultrathin
film adsorbed on a semi infinite substrate,
bounded by vacuum (or if we wish a capping layer)
on the other side. As we shall demonstrate below,
we can also calculate the linewidths of trilayers
adsorbed on a semi infinite substrate, so we can
study the linewidth of both the acoustic and
optical spin wave modes. Such structures have
been explored by the Baberschke group\cite{14}.
While, for reasons discussed below, it is
difficult to make a detailed comparison between
our results and their data, we do see features in
our results rather similar to those found
experimentally. Also, we obtain an excellent
account of the linewidths reported in Ref. [6]
for ultrathin films of Fe adsorbed on the Au(100)
surface. We have also examined other
film/substrate combinations with the aim of
exploring the influence of the electronic
structure of the substrate and the local atomic
geometry at the interface on the spin pumping
contribution to the linewidth.

The calculations reported here employ the
formalism and methodology we developed
earlier\cite{15} which was employed in our
successful quantitative account\cite{16}  of the
frequencies, linewidths and lineshapes associated
with large wave vector spin waves excited in spin
polarized electron energy loss (SPEELS) studies
of Co films adsorbed on the Cu(100)
surface\cite{17}. The very large linewidths found
in these experiments have their origin in the
same process which enters the spin pumping
contribution to the ferromagnetic resonance (FMR)
linewidth: in these itinerant materials, the
collective spin wave mode is damped by virtue of
the transfer of angular momentum to band
electrons. The spin pumping FMR linewidth then
may be viewed as the zero wave vector limit of
the large linewidths found in the SPEELS data. It
is the case, then, that this contribution
linewidth has a strong dependence on wave vector
parallel to the surface, for wave vector
variations on the scale of the surface Brillouin
zone.  It is also the case that our methodology
has led to quantitative accounts of both the
nature of the spin excitation spectrum\cite{18},
and the damping rates associated with these modes
throughout the entire surface Brillouin zone.

In section II of this paper, we discuss our
approach and comment on aspects of the numerical
computations. Our results are discussed in
section III, and section IV is devoted to
concluding remarks.

\section{Comments on the Analysis}

The approach we have employed has been discussed
in detail in our earlier
publications\cite{15,16},  so we summarize it
only briefly here. The most detailed discussion
is to be found in the first reference cited in
Ref. [15].  As remarked in section I, we employ
an empirical tight binding description of the
electronic structure of the various constituents
in the samples we model. Nine bands are included
for each material. These are the four sp bands
supplemented by the d band complex.
Ferromagnetism in the relevant films is driven by
intratomic, on site Coulomb interactions that
operate within the d shell. We refer to the model
Hamiltonian as a generalized Hubbard model. A
description of the ground state is generated in
mean field theory, allowing the magnitude of the
moments in each plane of each ferromagnetic film
to vary. In an early publication, we have shown
that our approach provides results for the
magnetic moment profile of the film and layer
dependent partial wave density of states in very
good accord with those generated by full density
functional calculations\cite{19}.

We then generate a description of the frequency
spectrum of spin fluctuations in the film through
calculating the appropriate wave vector frequency
dependent transverse susceptibility. This is done
through use of the Random Phase Approximation
(RPA) of many body theory, which is a
1``conserving approximation." The quantity we
generate by this means is written
$\chi_{+,-}(\vec{Q}_{\|},\Omega
;l_{\perp},l_{\perp}'$). It has the following physical
meaning. Suppose the magnetizations of the
various ferromagnetic films in the structure are
parallel to the $z$ axis. Then apply a circularly
polarized, space and time dependent field in the
$xy$ plane of the form
$\vec{h}(\vec{l}_{\|},l_{\perp}) = h(l_{\perp})\{
\hat{x}+i\hat{y}\}\exp
[i\vec{Q}_{\|}\cdot\vec{l}_{\|}-i\Omega t]$. Here
a lattice site in the structure is denoted by
$\vec{l} = \vec{l}_{\|}+\hat{n}l_{\perp}$, where
$\vec{l}_{\|}$ lies in the plane parallel to the
film surfaces, and $\hat{n}$ is a vector
perpendicular to the surfaces. The wave vector
$\vec{Q}_{\|}$  lies in the two dimensional
surface Brillouin zone of the structure, and
$\hat{x}, \hat{y}$ represent Cartesian unit
vectors in the $xy$ plane. In the present paper,
since we are interested in the ferromagnetic
resonance response, all calculations are confined
to zero wave vector, $\vec{Q}_{\|} \equiv 0$. The
applied magnetic field just described can have
arbitrary variation in the direction normal to
the film surfaces. If $\la
\vec{S}_+(\vec{l}_{\|},l_{\perp};t)\ra$  is the
expectation value of the transverse component of
spin at the site indicated, then linear response
theory provides us with the relation
\bqe
\la \vec{S}_+(\vec{l}_{\|},l_{\perp};t)\ra =
\left\{\sum\limits_{l_{\perp}'}\chi_{+,-}(\vQp
,\Omega; l_{\perp},l_{\perp}')\right\}\exp [i(\vQp
\cdot \vl_{\|}-i\Omega t)].
\eqe
Of central interest to us is the spectral density function
\bqe
A(\vQp ,\Omega ;l_{\perp}) = \frac{1}{\pi}Im \{
\chi_{+,-}(\vQp ,\Omega ;l_{\perp},l_{\perp '})\}
.
\eqe
It follows from the fluctuation dissipation
theorem that this quantity, considered as a
function of $\Omega$, provides us with the
frequency spectrum of spin fluctuations of wave
vector $\vQp$ on layer  $l_{\perp}$. Suppose for
the moment we consider a film of $N$ layers which
may be modeled by the Heisenberg Hamiltonian,
which envisions a localized spin situated on each
lattice site. Then for each choice of $\vQp$, we
have exactly $N$ spin wave modes, each with
infinite lifetime. Let $\Omega_a(\vQp )$ be the
frequency of one such mode, and let $e_a(\vQp
;l_{\perp})$ be the associated eigenvector. For
this model, which we note is inappropriate for
the itinerant electron ferromagnetic films
considered here\cite{15,16,18}, one may show that
\bqe
A(\vQp ,\Omega ;l_{\perp}) = \sum\limits_a \left|
e_a(\vQp ,l_{\perp})\right|^2 \delta (\Omega -
\Omega_a(\vQp )) .
\eqe
The form in Eq. (3) provides one with an
understanding of the information contained in the
spectral density function.

Of course, it would be highly desirable to base
the analysis of the spin dynamics of  systems
such as we explore through use of a state of the
art density functional description of the ground
state, combined with a time dependent density
functional analysis (TDDFA) of the spin
fluctuation spectrum. At the time of this
writing, one cannot carry out studies of the spin
fluctuation spectrum within the TDDFA for systems
as large as we have studied in the
past\cite{15,16} and that we explore in the
present paper. The demands on computation time
are prohibitive. We believe it is essential to
employ fully semi infinite substrates for proper
calculations of spin wave linewidths with origin
in decay of the collective excitations to the
Stoner excitation manifold. One must have a true
continuum of final states for a proper
description of the linewidth. All of our
calculations employ a full semi infinite
substrate, unless otherwise indicated. The virtue
of our multi band Hubbard model is that once the
irreducible particle hole propagator is computed,
inversion of the RPA integral equation is
straightforward and fast, by virtue of the fact
that the particle-hole vertex is separable in
momentum space within our framework. TDDFA
calculations employ an ab initio generated vertex
function which requires the numerical solution of
a full integral equation, once the irreducible
particle hole propagator is in hand. One can
handle only rather small systems within this
framework, at present.

If one describes our treatment of the ground state and
then the RPA description of the spin dynamics by
Feynman diagrams, it is the case that we include
the same set of diagrams as incorporated into the
full density functional theory, so in a certain
sense one may view what we do as a simplified
version of time dependent density functional
theory. However, technical simplifications such
as those described in the previous paragraph
enable us to address very large systems. We note
that our calculations involve no adjustable
parameters, since all of our input parameters are
taken from the appropriate literature. A rather
detailed discussion of how we proceed is
contained in Ref. [19], and considerable detail
is found in the first paper cited in Ref. [15].

In spite of the remarks above, it is the case
that the numerical work involved in the
calculation of the dynamical susceptibility is
extremely demanding from the computational point
of view. As discussed in detail in the first
paper cited in Ref. [15], to calculate the
irreducible particle hole propagator it is
necessary to calculate an energy integral whose
integrand is itself a two dimensional integral
over the full surface Brillouin zone. We have
also shown that very careful attention must be
paid to convergence of the Brillouin zone
integrations for physically reasonable results to
emerge from $\chi_{+,-}(\Omega
;l_{\perp},l_{\perp}')$. We refer the reader to
Fig. 6 of the first paper in Ref. [15] and the
associated discussion in the text. We find that
4224 points in the surface Brillouin zone must be
employed to insure proper convergence. The
irreducible particle hole propagator is a matrix
structure, whose size is the number of layers in
the film, N, times the number of orbitals
included in the calculation. Thus, for us to
analyze spin fluctuations in a 10 layer
ferromagnetic film, we must perform the Brillouin
zone integration just described for a $90\times
90$  matrix. For films deposited on semi infinite
substrates, of course we require the single
particle Green's function for the semi infinite
structure. This part of the computation also
consumes a considerable fraction of the total
computing time.

Fortunately, the problem in question can be
easily adapted to run in parallel on a cluster of
workstations. We adopted a very straightforward
parallelization strategy, in which we spread 64
points of energy integration among 32 processors.
For interprocess communication, we use the MPICH
implementation\cite{20} of the Message Passing
Interface (MPI)\cite{21}. The interprocess
communication overhead for this specific problem
is minimal, so that the execution time scales
roughly linearly with the number of processors.
This allows us to calculate the dynamic
susceptibility of relatively thick ferromagnetic
films ($\sim$ 15 layers) on a fully semi infinite
substrate, in a relatively short time. A
computation of the frequency response for one
film can be performed in less than 48 hours.
Typically to generate the absorption spectra
shown in section III, we use one hundred
frequencies.

\section{Results and Discussion}

In our previous studies of spin waves in various
ultrathin film/substrate combinations, our
emphasis was on excitations with relatively large
wave vectors. The excitation energies of such
modes are very large compared to the Zeeman
energy, which describes the interaction of the
spins in the system with externally applied
field. Thus the external field was taken to be
zero in all of our previous studies. Here, when
we examine the modes whose wave vector is
identically zero, with emphasis on the modes seen
in FMR experiments, quite clearly we must
incorporate the externally applied magnetic
field. In the results that follow, we have
immersed all the electrons in a spatially uniform
Zeeman field, and we denote the precession
frequency in this Zeeman field by $\Omega_0$.

We should say a few words about how the strength
of the Zeeman field has been chosen. First of
all, the energy scale used in our electronic
structure is the electron volt, or equivalently
the Rydberg. The spin excitations we studied in
our earlier papers had energies compatible with
this scale, in the range of 0.01 eV to 0.3 eV
\cite{15,16} depending on the wave vector we
chose to study. If we choose our applied Zeeman
field to be in the range actually used in
laboratory experiments, then the energy scale of
the FMR mode is so small ($\sim 10^{-5}$ eV) that
an enormous amount of computation time would be
required to produce calculations with an energy
grid so fine as to allow us to generate accurate
profiles for these very low energy modes. It is
the case, however, that so long as $\Omega_0$  is
very small compared to the typical electronic
energy scales, the linewidths scale linearly with
resonance frequency. We remark that we have
checked carefully that this is so, in our early
studies. So long as we remain in the regime where
the linewidth scales in this linear manner, we
may choose an unphysically large value for
$\Omega_0$, and scale the computed linewidths
appropriately when we compare our results with
data. What we choose to do instead is always to
plot the ratio $\Delta \Omega_0,\Omega_0$, with
$\Delta\Omega_0$ the linewidth determined as
discussed below. This is independent of applied
field, in the regime where the linearly scaling
holds. In the calculations to be presented in
this paper, we have chosen $\Omega_0 = 2\times
10^{-3}$ Rydbergs $=$27 meV.

In Fig. (1a), we show the absorption spectrum for
the FMR mode (the lowest lying spin wave mode
of the film, with wave vector parallel to the
film surface identically equal to zero) of a two
layer Co film adsorbed on the Cu(100) surface.
One sees the resonance line, with peak very close
to $\Omega_0$. In principle, there can be a $g$
shift associated with this mode even in the
absence of spin orbit coupling, since strictly
speaking the amplitude of the spin motion in the
direction normal to the surface is not perfectly
uniform. In our earlier studies of the electron
spin resonance response of isolated magnetic
adatoms adsorbed on the Cu(100) surface, we found
such $g$ shifts to be quite large. Here we find
that the peak of the absorption line coincides
exactly with $\Omega_0$, so far as we can tell.

There is another idealization we have made that
must be mentioned. In an actual sample, where a
ferromagnetic film is on top of a nonmagnetic
substrate such as Cu, when the spins precess in
the ferromagnet, dynamic dipolar fields are
generated by the spin motion. Thus, for a film
magnetized parallel to its surfaces, the FMR
frequency is shifted from $\Omega_0 = \gamma H_0$
to $\gamma [H_0(H_0 +4\pi M_s)]\sfr$, where
$\gamma$ is the gyromagnetic ratio for the
ferromagnetic film, $H_0$ is the applied Zeeman
field, and $M_s$ is the saturation magnetization.
It is also the case that the gyromagnetic ratio
for the ferromagnetic film will differ from that
of the conduction electrons in the substrate.
Thus, in an actual sample, the precession
frequency of the spins in the ultrathin
ferromagnetic film will differ from that for
electrons in the substrate. We have explored the
influence of this effect by varying the ratio of
the Zeeman frequency of the spins in the
ferromagnet, to that of the spins in the
substrate by as much as a factor of two.  We find
that our linewidths, normalized to the resonance
frequency in the ferromagnet, are insensitive to
this difference. On physical grounds one expects
this. In microscopic language, the linewidths we
calculate have their origin on the transfer of
angular momentum from the coherently precessing
moments in the ferromagnet to the bath of
itinerant band electrons. If we think of this as
a form of Landau damping, where the spin wave
decays to the bath of Stoner excitations of the
film/substrate complex (spin triplet particle
hole excitations), a shift in the precession
frequency of the ferromagnetic film relative to
that of the substrate electrons is compensated by
a very tiny shift in the wave vector transfer
involved in the decay process, on the scale of
the relevant two dimensional Brillouin zone. Note
that wave vector components normal to the surface
are not conserved for the systems we study. Thus,
in Fig. 1(a) and in the results to be discussed
below, all electrons in the system are subjected
to the same external field, and are assumed to
have the same gyromagnetic ratio, with dipolar
fields in the ferromagnet ignored.

In Fig. 1(b), we show calculations of the of a
Co$_2$Cu$_2$Co$_2$ trilayer adsorbed on a semi
infinite Cu(100) surface. One now sees two modes,
the acoustical and the optical spin wave mode of
the ferromagnetic bilayer. It is clear from
visual inspection that the acoustical mode is
narrower than the FMR line shown in Fig 1(a),
while the optical mode is substantially broader,
a pattern seen in experimental studies of such
systems\cite{14}. It should be noted that in Fig.
1(b), it is not quite the FMR absorption spectrum
that is displayed. For two such identical films,
the antisymmetric optical spin wave mode has no
net transverse magnetic moment, and thus would be
silent in an FMR spectrum. In actual FMR
experiments where both modes are observed, the
two ferromagnetic films are inequivalent. What we
display in Fig. 1(b) is the spectral density
function $A(\vQp = 0, \Omega ;l)$,
where we have taken $l_{\perp} = 1$, where the 1
refers to the outermost layer of the outermost
film. The physical interpretation of this
spectral density function is that it describes
the frequency spectrum of the fluctuations of the
spins in the atomic layer indicated, at zero wave
vector parallel to the surface. The acoustic and
optical spin wave both leave their signature on
this response function.

One final remark is that our procedure for
determining the linewidths discussed below is as
follows. For each structure of interest, we
calculate a spectrum such as that illustrated in
Fig. 1(a) and Fig. 1(b). Then we fit the curves
to appropriate Lorentizians, and from this we
extract a linewidth. The quantity $\Delta\Omega$
is the half width at half maximum.

The first case we discuss is that of Fe films
adsorbed on the Au(100) surface. This is the
system studied by Urban, Woltersdorf and
Heinrich\cite{6}.

In Fig. 2, we show our calculated linewidths as
open circles, for Fe film thicknesses in the
range of two to ten monolayers. The solid line is
a best power law fit to the data, which turns out
to be 1/(N$_{\rm Fe})^{0.98}$ law, with N$_{\rm
Fe}$ the number of Fe layers. Various
authors\cite{7,8,12} have argued that the spin
pumping linewidth should fall off inversely with
the thickness of the ferromagnetic film, so these
result are consistent with this picture. Our
calculations show this behavior nicely. For small
thicknesses, we see oscillations of modest
amplitude around the 1/N$_{\rm Fe}$ law.  These
are quantum oscillations such as those discussed
in Ref. [10], though as discussed by the authors
of Ref. [13] the simple quantum well model
exaggerates their amplitude.

The solid circles in Fig. 2 are data taken from
Ref. [6]. Theory and experiment agree very
nicely. Unfortunately it is difficult for us to
carry out calculations for films much thicker
than ten layers for the FMR mode. In our
integrations over energy, a small imaginary part
is added to energy denominators, and we must
always keep this small compared to the linewidth.
We must decrease this as the film gets thicker,
and at the same time we must increase the density
of points in our integration grids to generate a
reliable, converged line shape uninfluenced by
the numerical procedures.

It is interesting to inquire how the linewidth
depends on the electronic structure of the
substrate. To explore this, we have carried out
calculations for anFe film deposited on the
W(110) surface. While we are unaware of any FMR
studies of this system, the magnetism associated
with the Fe/W(110) system has been extensively
studied in the
literature.  Our results are given in Fig. 3. The
straight line is a best fit to a power law
dependence on the Fe film thickness, and in this
case we find 1/(N$_{\rm Fe})^{0.81}$, just a bit
slower that a simple variation inversely
proportional to the film thickness.  However, we
again see quantum oscillations about a simple
power law fit, at smaller thickness, so a simple
power law is a bit of an oversimplification.
Surprisingly, in our view, the numerical values
of the linewidths displayed in Fig. 3 are not so
very different than those in Fig. 2. We had
expected larger linewidths, because of the large
density of states in the substrate associated
with the unfilled d bands of W. It should be
pointed out, however, that the connectivity of an
Fe film is larger when it is adsorbed on Au(100)
than when it is adsorbed on W(110). An Fe atom at
the interface with the Au(100) surface is
connected to four nearest neighbor Au atoms,
whereas it is coupled to just two nearest
neighbor W atoms in the Fe/W(110) interface. One
virtue of our empirical tight binding approach is
that we can probe the physics responsible for
results such as these by artificially selecting
parameters. We turn to such studies next, to
obtain insight into this result.

It is interesting to compare the calculated
linewidths for an Fe film adsorbed on W(110), and
for an identical film adsorbed on a bcc Cu
crystal, which we can simulate with our approach.
Here the connectivity to the substrates is the
same. Of course, in the first case the Fermi
level intersects the d band complex of the
substrate, whereas in the second case the d bands
are well below the Fermi level. For the cases of
a two layer Fe film on W(110) we see from Fig. 3
that $\Delta\Omega /\Omega_0$ assumes the value
$1.8 \times 10^{-2}$, whereas for the Fe film on
Cu(110), the line is actually a bit broader, with
$\Delta\Omega /\Omega_0 = 2.7\times 10^{-2}$.
Even though the two numbers are rather close to
each other, it would seem that the physics which
underlies the transfer of spin angular momentum
across the interface is rather different in the
two cases. In Fig. 4, we show calculations which
illustrate this. Let us first look at the right
panel, the case of Fe on W(110). The solid curve
is the spectral density calculated with use of
the full electronic structure. The dashed curve
is a calculation in which the d-d hopping terms
are set to zero across the interface. The line
narrows very substantially when the d-d hopping
is shut off, as we see. Evidently for this case,
direct communication between the d electrons in
the Fe film and the W substrate plays a central
role in the transfer of spin angular momentum.
The electrons at the W Fermi surface have strong
d character, and the electrons in the Fe with
strong sp character play a minor role in the
damping process. This view is reinforced by the
dot dash curve in the right hand panel of Fig. 4,
where the sp hopping terms are set to zero across
the interface. We see little change in the
spectral density. The situation is very different
for the case of Fe on Cu(110), as we see from the
left hand panel in Fig. 4. We see here that
communication between both the sp and the d like
portions of the electronic wave function across
the interface play an important role. Even though
the 3d bands of Cu are completely filled, there
is d character admixed in the electron wave
functions in the vicinity of the Fermi energy,
while at the same time coupling through the sp
character of the wave function enters as well.
Thus, while the final numbers for the spin
pumping contributions to the linewidth are rather
similar for these two very different substrates,
the physical picture which underlies the transfer
of spin angular momentum across the interface is
rather different. In view of this discussion, it
would be of great interest to see experimental
FMR linewidth studies for the case of Fe on
W(110).

We turn next to a discussion of our calculation
of linewidths of the FMR modes of
ferromagnetic/nonmagnetic metal/ferromagnetic
film combinations adsorbed on semi infinite
substrates. We have already seen that such
structures produce two features in the absorption
spectrum, one from an acoustic spin wave mode,
and one from an optical mode. The splitting
between these two modes is controlled by the well
known interfilm exchange coupling transmitted
through the non magnetic spacer layer. As
remarked above, if one has a sample with two
identical ferromagnetic films, only the
acoustical mode is active in an FMR experiment,
since the optical mode possesses no
net transverse magnetic moment. In samples
fabricated from ultrathin films, the thickness of
the trilayer is small compared to the microwave
skin depth, so to excellent approximation the
films are illuminated by a spatially uniform
magnetic field. We note that in Brillouin light
scattering, one may observe both modes, since the
optical skin depth is in the range of 10-20 nm
for materials of current interest. Here, the
exciting optical field is spatially non uniform,
on the length scale of the structure. It is the
case, though, that in any real sample, the two
films will always be inequivalent even if their
thicknesses are the same. For instance, both
interfaces of the innermost film are in contact
with non magnetic metals, with the substrate on
one side and the spacer layer on the other. The
outer film has vacuum on one side, and the spacer
layer on the other. Hence, the anisotropy fields
will be different in each film, and this will
render them inequivalent. As discussed below, we
have simulated such effects by applying different
external magnetic fields to each film. First,
however, we present our results for the case
where the two ferromagnetic films are regarded as
identical in character.

Before we turn to our results, we remark on how
we proceed with our determination of the
linewidths for the optical and the acoustical
mode of the trilayer structure. In Fig. 1(b), we
show the spectral density function for a case
where the Cu spacer layer is sufficiently thin
that the two modes can be resolved easily in the
spectral density associated with one selected
layer. As the Cu spacer is made thicker, the
splitting between the two modes decreases and if
we plot the spectral function illustrated in Fig
1(b), we simply see a single asymmetric line,
with the optical mode buried in the wing of the
acoustical mode. We have proceeded as follows.
For each frequency, we form a two by two matrix
with elements $\chi_{11} =
\sum\limits_{l_{\perp},l_{\perp}'\epsilon
1}Im\{\chi_{+,-}(0,\Omega ;l_{\perp},l'_{\perp})$,
$\chi_{12} = \sum\limits_{l_{\perp}\epsilon
1,l'_{\perp}\epsilon 2}Im\{ \chi_{+,-}(0,\Omega
;l_{\perp},l'_{\perp})\}$ and so on. In the
definition of $\chi_{11}$, the sums on both
$l_{\perp}$ and $l_{\perp}'$ range over the
atomic planes in the outermost film, while in
$\chi_{12}$ the sum on $l_{\perp}$ ranges over
the atomic planes of the outermost film, and the
sum over $l'_{\perp}$ ranges over the atomic
planes of the innermost film. Diagonalization of
this matrix at each frequency leads us to two
eigenvectors, one with acoustical character and
one with optical character. If we generate plots
of the spectral densities of these two
characteristic motions, in one we see only the
acoustical feature, and in the second we see only
the optical feature. We illustrate this in Fig.
5.  In Fig. 5(a) we show the spectral density
function associated with spin fluctuations in the
innermost layer of the inner film of a
Co$_2$Cu$_4$Co$_2$/Cu(100) structure. The optical
mode appears in the wing of the acoustical mode,
and one cannot reliably extract a width for the
mode from this response function. In Fig. 5(b) we
see how one may isolate the acoustical mode by
the procedure just describe, and in Fig. 5(c) we
show the optical mode.

We remark that for the samples studied
experimentally in Ref. [3], the two ferromagnetic
films in the trilayer have very different
resonance frequencies when taken in isolation, so
the issue of discriminating between the two modes
did not arise. Save for one measurement discussed
below, the trilayer structures studied in Ref.
[3] may be viewed as two oscillators whose
resonance frequencies differ by an amount much
larger than their linewidths, coupled weakly by
the interfilm exchange coupling mediated by the
spacer layer.

In our earlier studies of large wave vector spin
waves and their dispersion\cite{16} we have
compared the dispersion relation which emerges
from our dynamical theory with that calculated on
the basis of adiabatic theory, where exchange
couplings between spins are calculated within
quasi static theory, and a Heisenberg Hamiltonian
constructed from these is used to generate a spin
wave dispersion curve. We found substantial
differences between the two cases. The strong
damping present at large wave vectors leads to
shifts in the frequency of the spin wave. This
effect, which softens the modes, is not contained
in calculations based on an adiabatic description
of energy changes associated with rotations of
the moments. One can inquire if the effective
interfilm exchange couplings deduced from the
splitting between the acoustical and optical spin
wave modes of the trilayer are the same as those
described by adiabatic theory.

We provide a comparison which addresses this
issue in Fig. 6. The solid line shows the
dependence of the interfilm coupling strength on
$N$, the number of Cu layers between the two
Co$_2$ films, calculated adiabatically using the
method in Ref. [23]. The dashed line provides the
results deduced from the splitting between the
two spin wave modes of the structure. The two
agree rather well, though there are indeed small
differences between the two.

It is evident that we find the wide of the
optical mode of the trilayer significantly larger
than that of the acoustic mode. This is
consistent with the results reported in Ref.
[14], as noted earlier. Very interesting to us is
the data in Fig. (6) of Ref. [14], where the
linewidth measured for these two modes is given
as a function of the number of Cu layers between
the two ferromagnetic films. One sees very clear
quantum oscillations in these two linewidths. In
our Fig. 7, we show calculations of the variation
of the acoustic and optic mode linewidths with
$N$, the number of Cu spacer layers in our
Co$_2$Cu$_N$Co$_2$/Cu(100) structure. The
calculations reproduce the features found in the
data strikingly well. For instance, in our
calculations we see a peak in the linewidth of
the acoustic mode at six atomic layers of Cu and
a second somewhat smaller peak at nine layers.
Both features appear in the data, indeed with the
peak at six layers the most prominent, Also very
much as in the data, for the optical mode we see
the peak at six layers in our calculations, while
the peak at nine layers is suppressed strongly.
In our calculations, the optical mode linewidth,
reckoned relative to that of the acoustic mode,
is larger than that in the data. At the peak at
six layers, we find the width of the optical mode
larger than that of the acoustical mode by a
factor of roughly five, whereas in the data this
ratio is a bit less than a factor of two. A
direct comparison of this ratio with the data is
not so relevant in our minds, since the resonance
frequencies of our two Co$_2$ films (each taken
in isolation) are identical, whereas that of the
Co$_{1.8}$ film and the Ni$_9$ film used in Ref.
[14] are different.

Some of the data in Ref. [14] has been taken on
Ni$_8$Cu$_N$Ni$_9$/Cu(100) samples. The resonance
frequencies of the Ni$_8$ and Ni$_9$ films differ
substantially when the externally applied field
is in plane, presumably because of the near
proximity of these Ni films to the spin
reorientation transition. However, the authors of
Ref. [14] were able to sweep the resonance
frequency of one film through the second by
varying the angle of the extermal applied
magnetic field with respect to the plane of the
films. They argue that the linewidths of the two
modes approach each other when the resonance
frequencies coincide. This is illustrated in
their Fig. 4, for a sample with $N=12$. For such
a thick copper spacer layer, the interfilm
exchange coupling is negligible.

We can simulate this with our
Co$_2$Cu$_N$Co$_2$/Cu(100) structures by applying
a different magnetic field to each film. In Fig.
8 we present results of a study where the
resonance frequency of the outer film is
$\Omega_0$ and that of the inner film is
$\lambda\Omega_0$, where $\lambda$ is varied from
1 to 2. By varying $\lambda$, we can study how
the linewidth of each mode behaves if the two
films initially are detuned $(\lambda = 2)$, and
then brought to the point where the isolated film
resonance frequencies coincide $(\lambda = 1)$.
Our calculations are carried out for $N=10$,
where the interfilm exchange coupling is very
weak.

We presume the authors of Ref. [14] have fitted
their FMR spectra to a sum of two Lorentizians in
order to extract the width of the two modes. We
have attempted a similar procedure with
simulations of the FMR absorption spectrum of the
structure. However, we find that when $\lambda$
is fairly close to unity, we cannot reliably
extract linewidths for the two modes by fitting
the overall spectral density in this manner. In
Fig. 8(a), the solid line shows the absorption
spectrum of the trilayer for the case where
$\lambda = 1.2$, and we see the high frequency
mode as a barely perceptible feature on the high
frequency side of the line.  The solid line is
the function $a(\Omega ) =
\sum\limits^{4}_{l_{\perp}'=1}\sum\limits^{4}_{l_{\perp}=1}Im\{
\chi_{+,-}(0,\Omega ;l_{\perp},l_{\perp}')\}$.
This gives the absorption spectrum of the
trilayer, if it is illuminated by a spatially
uniform microwave field. Also in Fig, 8(a), we
give the absorption spectrum for the outer film
by displaying $A_1(\Omega )$, defined in a manner
similar to $A(\Omega )$, except the spatial sums
are now confined to the outer film. Similarly,
$A_2(\Omega )$  is the absorption spectrum of the
inner film.  We have found that we can only
extract linewidths meaningfully from our
simulations if we isolate the optical and
acoustical modes by the procedure outlined
earlier in this section. Of course, this cannot
be done from experimental FMR spectra. In Fig.
8(b), we show the variation of the linewidths of
the two modes with $\lambda$. We see that
$\lambda$   is decreased from the value 2 toward
unity, the optical mode linewidth indeed
approaches that of the acoustical mode, but then
when $\lambda$  comes close to 1, the linewidths
diverge. The acoustic mode narrows, and the
optical mode broadens, as we have seen for the
calculations presented earlier where $\lambda$
was taken to be unity.

\section{Final Comments}

In this paper, through analysis of spectral
densities calculated at zero wave vector, we have
extracted linewidths of the FMR modes of the
structures we have explored. The origin of these
linewidths, which are built into the theoretical
methodology we have presented in the earlier
paper, is in the spin pumping mechanism discussed
elsewhere in the literature. We have seen that we
obtain an excellent account of the linewidths
reported in Ref. [6] for Fe films on Au(100), and
we also obtain a very good description of the
quantum oscillations in linewidth reported in
Ref. [14], for both the acoustical and optical
modes of ferromagnetic trilayers grown on the
Cu(100) surface.

These calculations employ the same theoretical
methodology that we have developed and applied
earlier to the analysis of the SPLEED data
reported in Ref. [17]. In the SPLEED
measurements, of course, it is large wave vector
spin excitations that are studied. In  Ref. [16],
we obtained an excellent account of the
dispersion of the single, broad asymmetric
feature seen in SPEELS, along with an excellent
account of both its width and asymmetric
lineshape. Thus, the calculations presented in
the present paper show that our method accounts
very nicely for the damping of spin motions in
these ultrathin films throughout the two
dimensional Brillouin zone, from the zero wave
vector excitations explored in the FMR
experiments to the large wave vectors probed by
SPEELS.

It is interesting to offer a physical viewpoint
of the FMR linewidths calculated here that is
somewhat different (but complementary to) that
set forward in papers devoted specifically to FMR
linewidths. First, suppose we were to consider
the FMR linewidth in an infinitely extended,
three dimensional crystal described by a model
such as we use here, in which electron energy
bands result from interactions of electrons with
the crystal potential, and ferromagnetism is
driven by Coulomb interactions between the
electrons. The FMR mode is the infinite
wavelength, or zero wave vector mode of the
system. Since such a Hamiltionian is form
invariant under rigid rotations of the spins, the
Goldstone theorem require the linewidth of the
zero wave vector mode to be identically zero.
This is a rigorous statement, independent of
approximations used in any specific calculation
to study spin waves. Our mean field description
of the ground state combined with the RPA
description of the spin dynamics, is a conserving
approximation and thus all of our conclusions are
compatible with the Goldstone theorem. At zero
field we will then find a width of identically
zero, for the zero wave vector spin wave of the
infinitely extended crystal. In real materials,
of course, there is a finite width to the FMR
mode, described phenomenologically by the damping
term in the well known Landau Lifschitz Gilbert
equation.   The origin of the linewidth in real
materials is the spin orbit interaction, since
its introduction into the Hamiltonian produces a
structure no longer invariant under rigid body
rotations of all spins in the system.

Now when we place a thin ferromagnetic film on a
non magnetic substrate, as in the systems
considered here, translational symmetry normal to
the surface is broken, and only the components of
wave vector parallel to the surface remain good
quantum numbers. The lowest lying mode with $\vQp
= 0$ is the FMR mode. This is not a uniform mode
of the system, by virtue of the breakdown of
translational symmetry normal to the surfaces;
the amplitude of the spin motion in the non
magnetic substrate differs from that in the
ferromagnetic film, for instance. Hence, even in
the absence of spin orbit coupling, one realizes
a finite linewidth for this mode. This is a way
of viewing the ``spin pumping" contribution to
the linewidth. Now in the limit that the
thickness of the ferromagnetic film becomes
infinite, the FMR mode of the film must evolve
into the uniform mode of the infinite crystal
whose linewidth must vanish in the absence of
spin orbit coupling. Hence the spin pumping
contribution to the linewidth falls to zero, with
increasing ferromagnetic film thickness.

Other questions will be interesting to explore.
Our method will allow us to explore the wave
vector dependence of the linewidth, for example.
Studies of this issue and other questions are
underway presently.

\acknowledgments

During the course of this investigation, we have
enjoyed stimulating conversations with Prof. B.
Heinrich and with Prof. K. Baberschke. This
research was supported by the U. S. Department of
Energy, through Grant No. DE-FG03-84ER-45083.
A.T. C. and R. B. M. also received support from
the CNPq, Brazil. A.T.C. also acknowledges the
use of computational facilities of the Laboratory
for Scientific Computation/UFLA.

\newpage
\begin{figure}
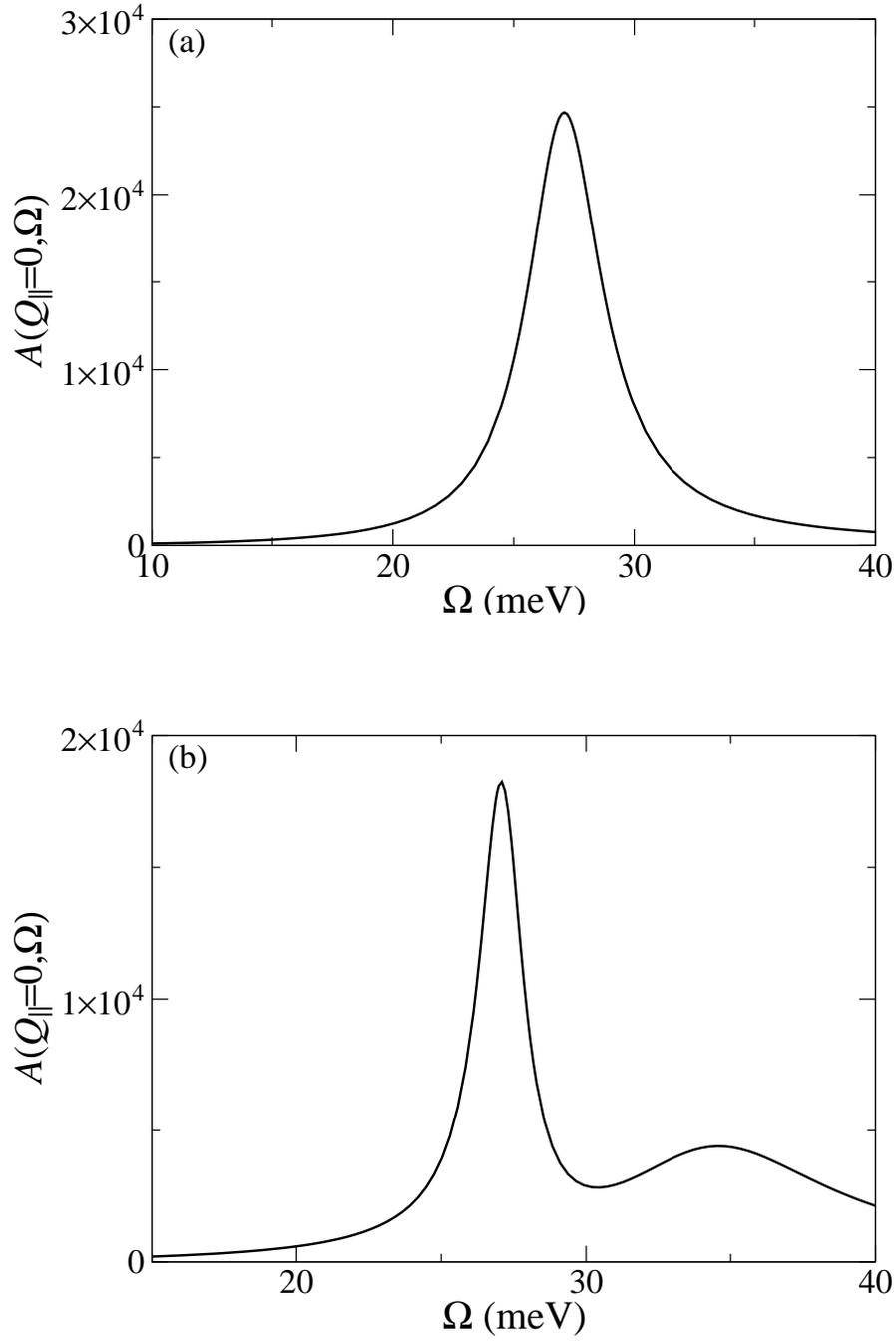

\begin{center}
\epsfig{file=fig1a.eps,scale=.5}\vspace*{.5in}\\
\epsfig{file=fig1b.eps,scale=.5}
\end{center}
\caption{We plot the spectral density function
$A(\vQp = 0, \Omega ;l_{\perp}$ for two cases:
(a) a two layer Co film adsorbed on the Cu(100)
surface, and (b) a trilayer consisting of two Co
films separated by two layers of Cu, with the
complex adsorbed on the Cu(100) surface. Each of
the Co films has two layers. The Zeeman energy
has been taken to be $\Omega_0 = 2\times 10^{-3}$
Rydbergs.   Here $l_{\perp}$ is chosen to be the
innermost atomic layer of the magnetic structure.}
\end{figure}
\begin{figure}
\begin{center}
\epsfig{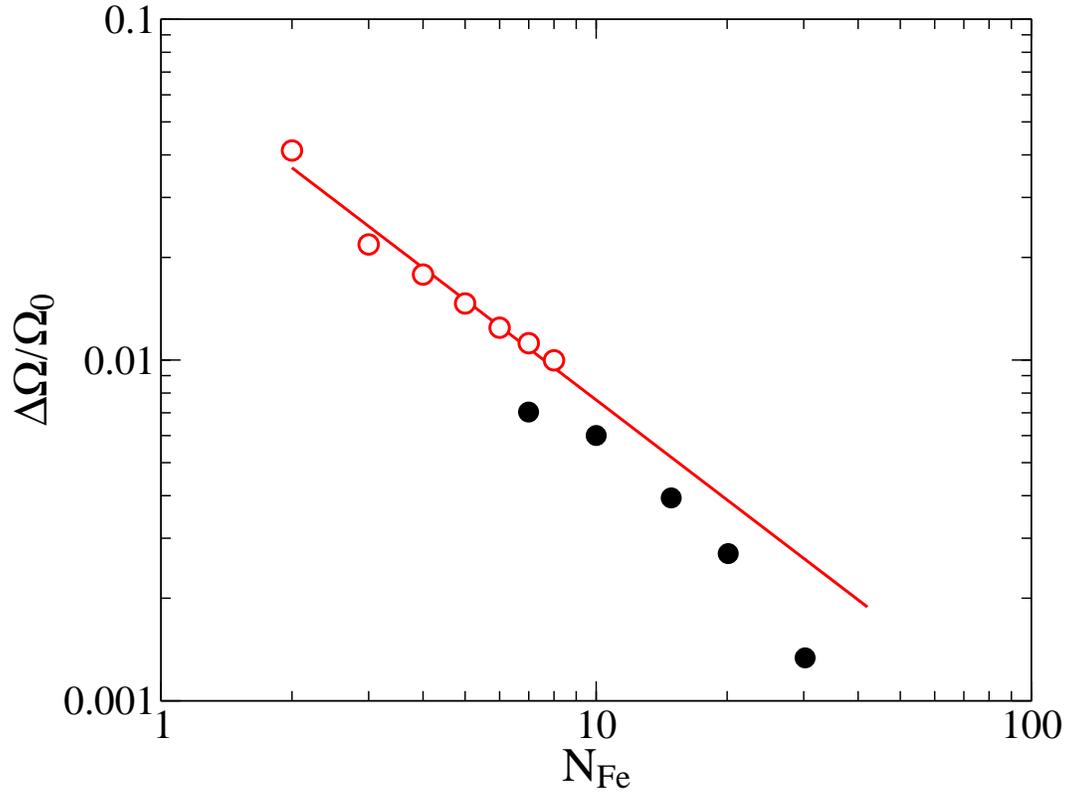}
\end{center}
\caption{For an Fe film on the Au(100) substrate,
we plot the ration $\Delta \Omega /\Omega_0$ as a
function of N$_{\rm Fe}$, the number of Fe
layers. The results of the calculations are given
as open circles, and the solid line is the best
power law fit to the data, which is 1/(N$_{\rm
Fe})^{0.98}$. The solid circles are taken from
measurements reported in Ref. [6].}
\end{figure}
\begin{figure}
\begin{center}
\epsfig{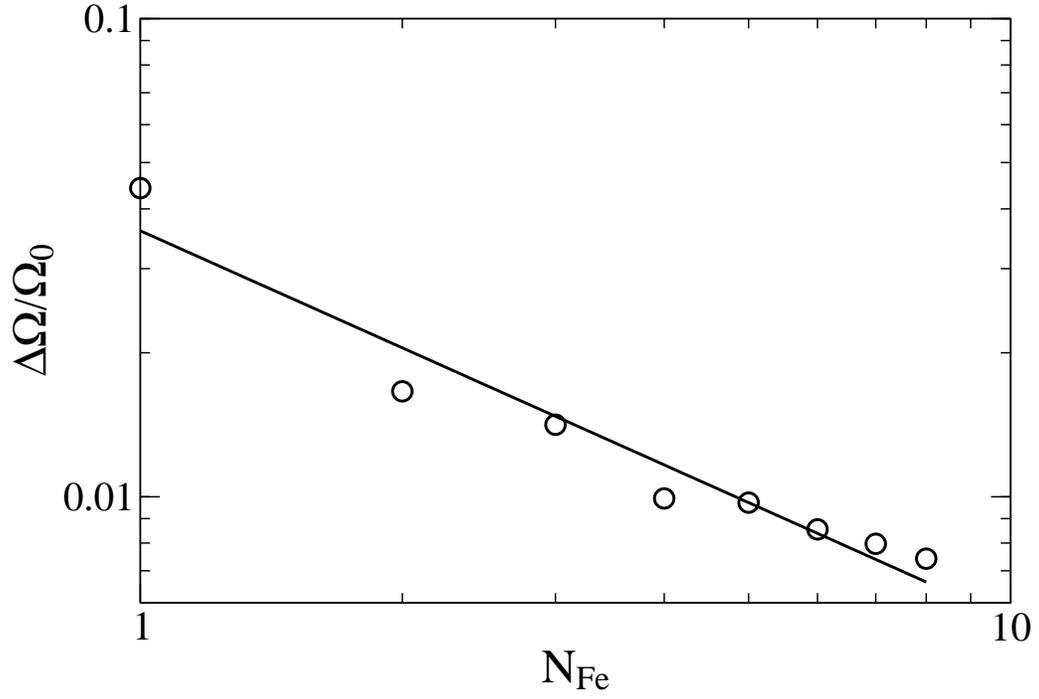}
\end{center}
\caption{The ratio $\Delta\Omega /\Omega_0$, for
an Fe film adsorbed on the W(110) surface. The
straight line is a best power law fit to the
calculations, and gives a fall off with film
thickness of 1/(N$_{\rm Fe})^{0.81}$. Clearly
quantum oscillations such as discussed in  Ref.
[10] are present for small film thicknesses, so
the power law fit is an oversimplification.}
\end{figure}
\begin{figure}
\begin{center}
\epsfig{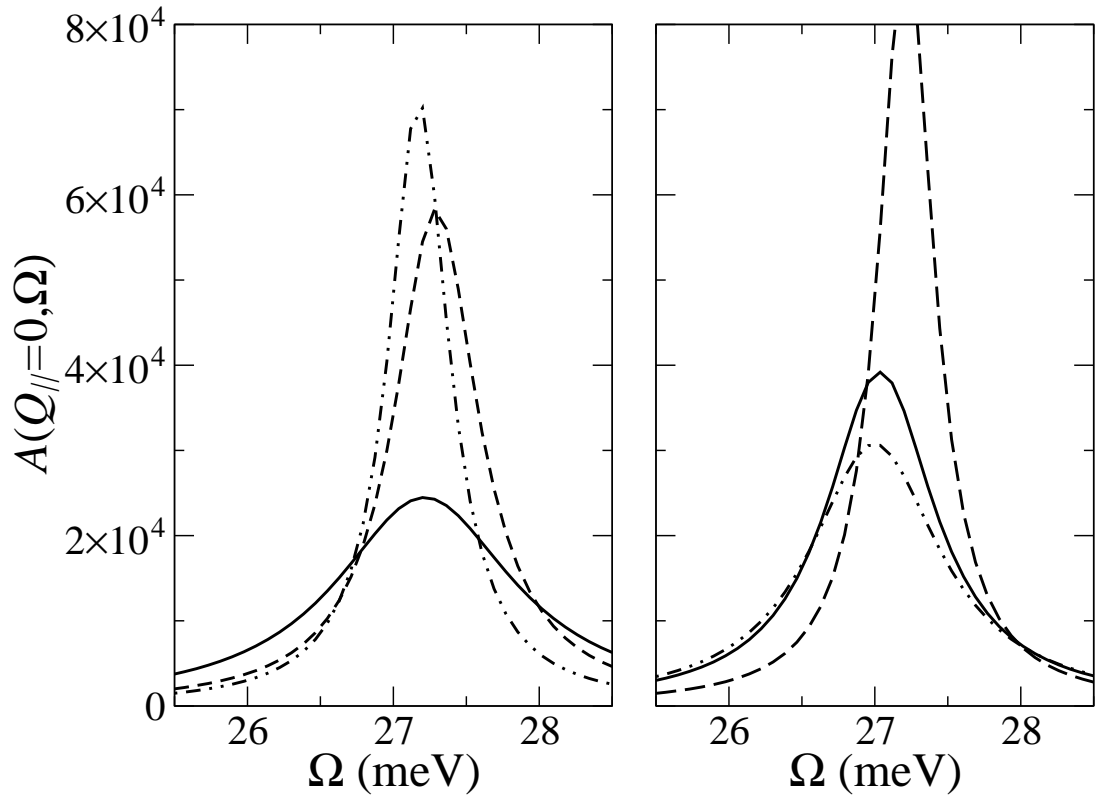}
\end{center}
\caption{For a two layer Fe film on Cu(110) (left
panel) and for such a film on W(110) (right
panel), we show how the linewidth is influenced
by various coupling across the interface. In both
cases, the solid curve shows the spectral density
function generated by a complete calculation. The
dashed curve has the d-d hopping terms across the
interface set to zero, whereas the dot dash curve
has the sp-sp hopping terms turned off.}
\end{figure}
\begin{figure}
\begin{center}
\epsfig{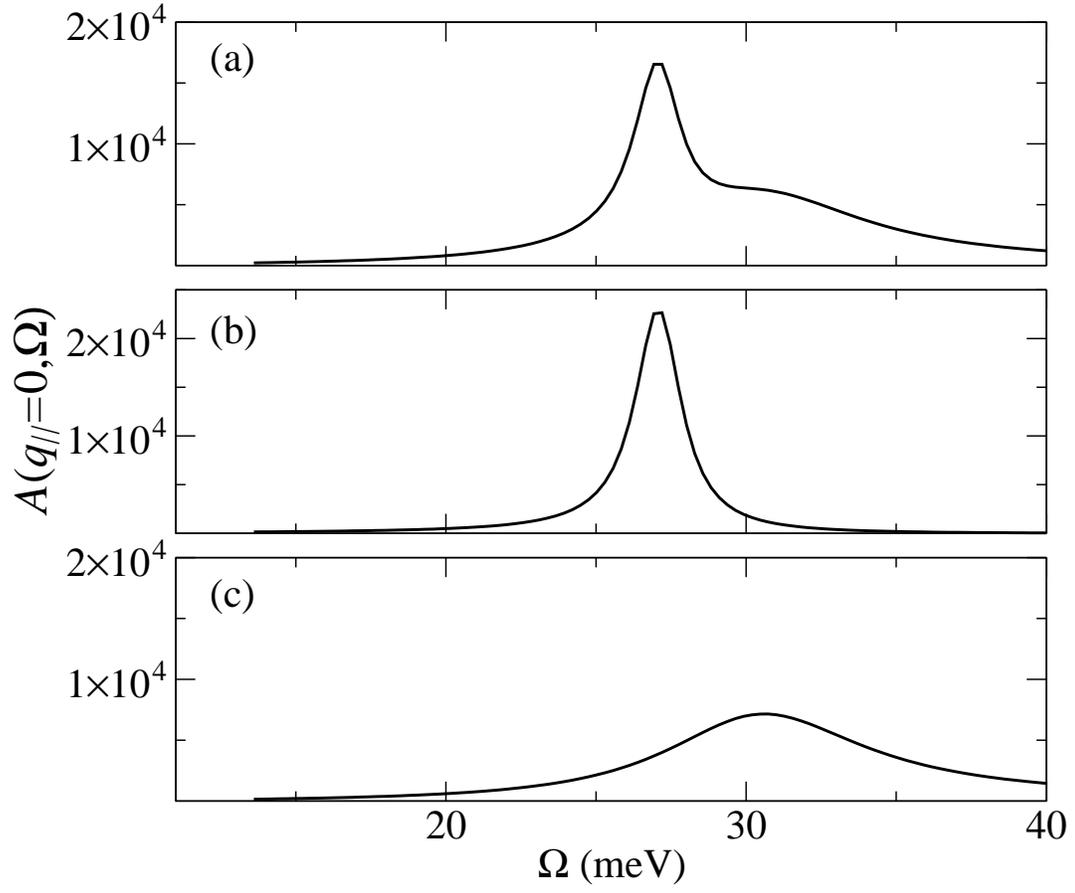}
\end{center}
\caption{For a Co$_2$Cu$_4$Co$_2$/Cu(100)
structure, we show (a) the spectral density
function $A(0,\Omega;l_{perp})$, for the case
where $l_{\perp}$  is chosen to be the innermost
layer of the inner film. Then in (b) we show the
acoustical mode profile, where the mode has been
isolated by the procedure described in the text.
In (c), we display the optical mode.}
\end{figure}
\begin{figure}
\begin{center}
\epsfig{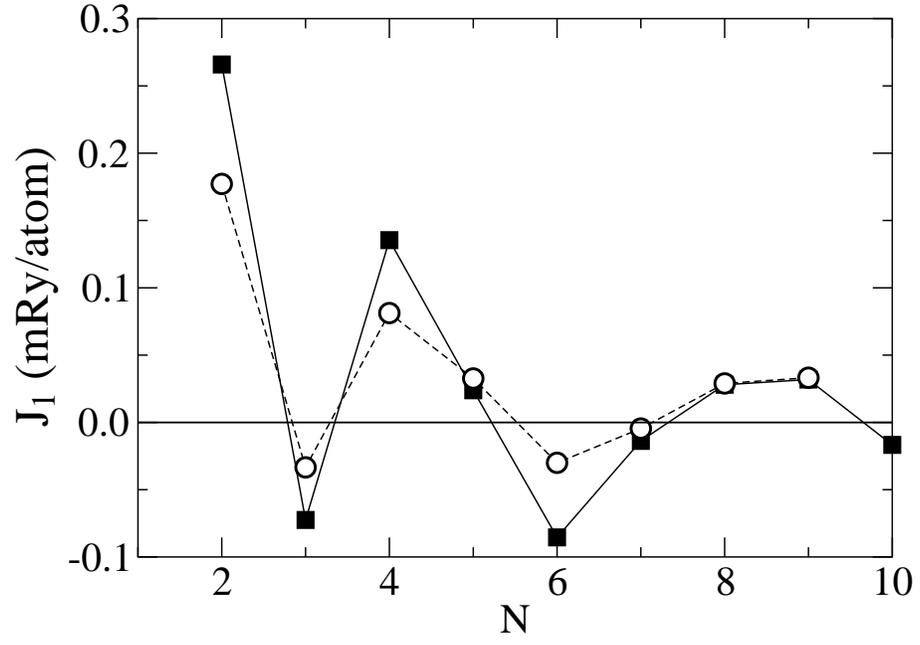}
\end{center}
\caption{For a Co$_2$Cu$_N$Co$_2$/Cu(100)
structure, we compare interfilm couplings deduced
from our dynamic calculations of the splitting
between the optical and acoustic mode of the
trilayer (open circles, dashed line) with that
calculated by adiabatic theory (solid circles,
solid line).}
\end{figure}
\begin{figure}
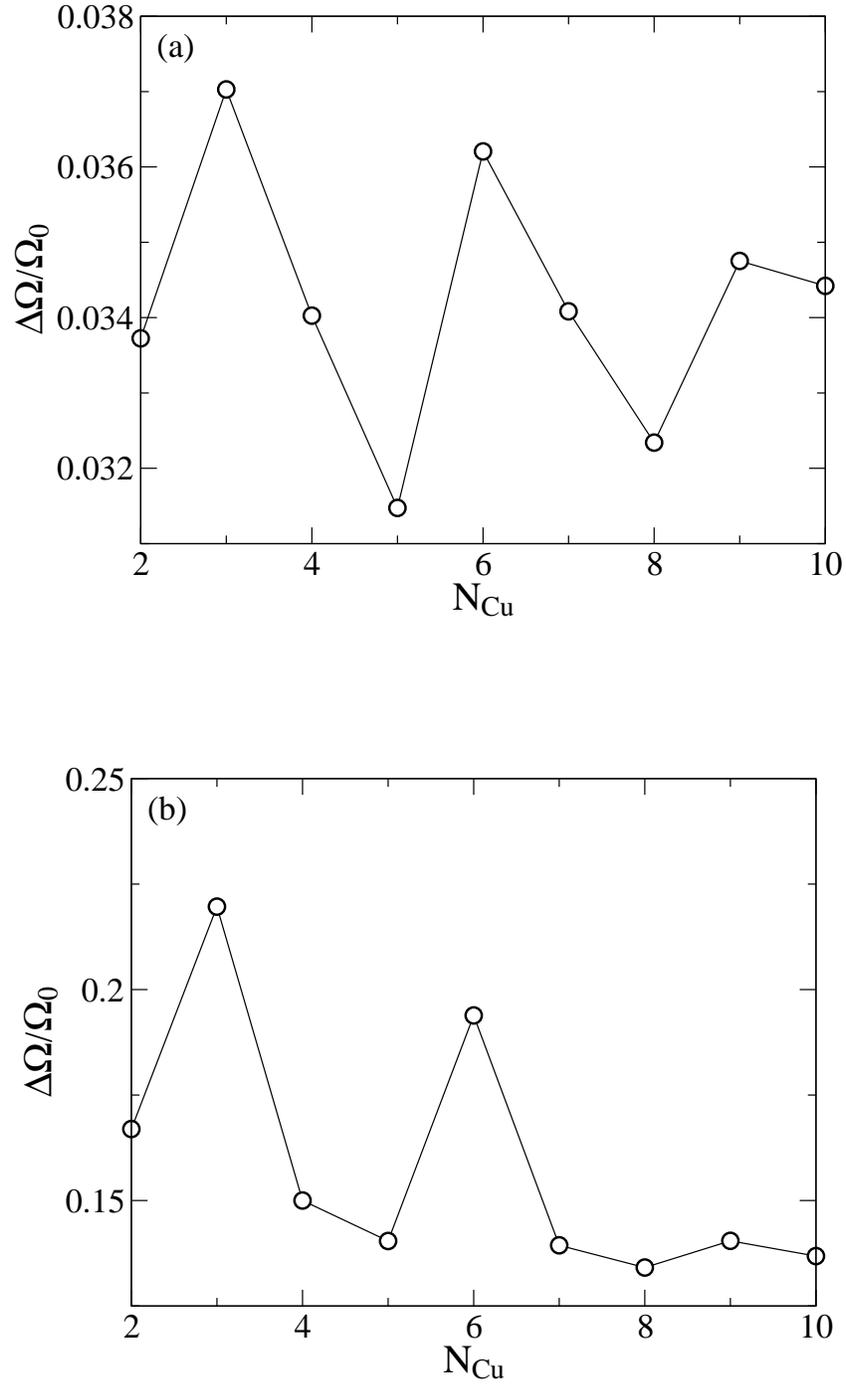

\begin{center}
\epsfig{file=fig7a.eps,scale=.5}\vspace*{.75in}\\
\epsfig{file=fig7b.eps,scale=.5}
\end{center}
\caption{For a Co$_2$Cu$_N$Co$_2$/Cu(100)
trilayer, we show the variation with $N$ of the
linewidth for (a) the acoustic spin wave mode of
the structure, and (b) the optical spin wave mode
of the structure.}
\end{figure}
\begin{figure}
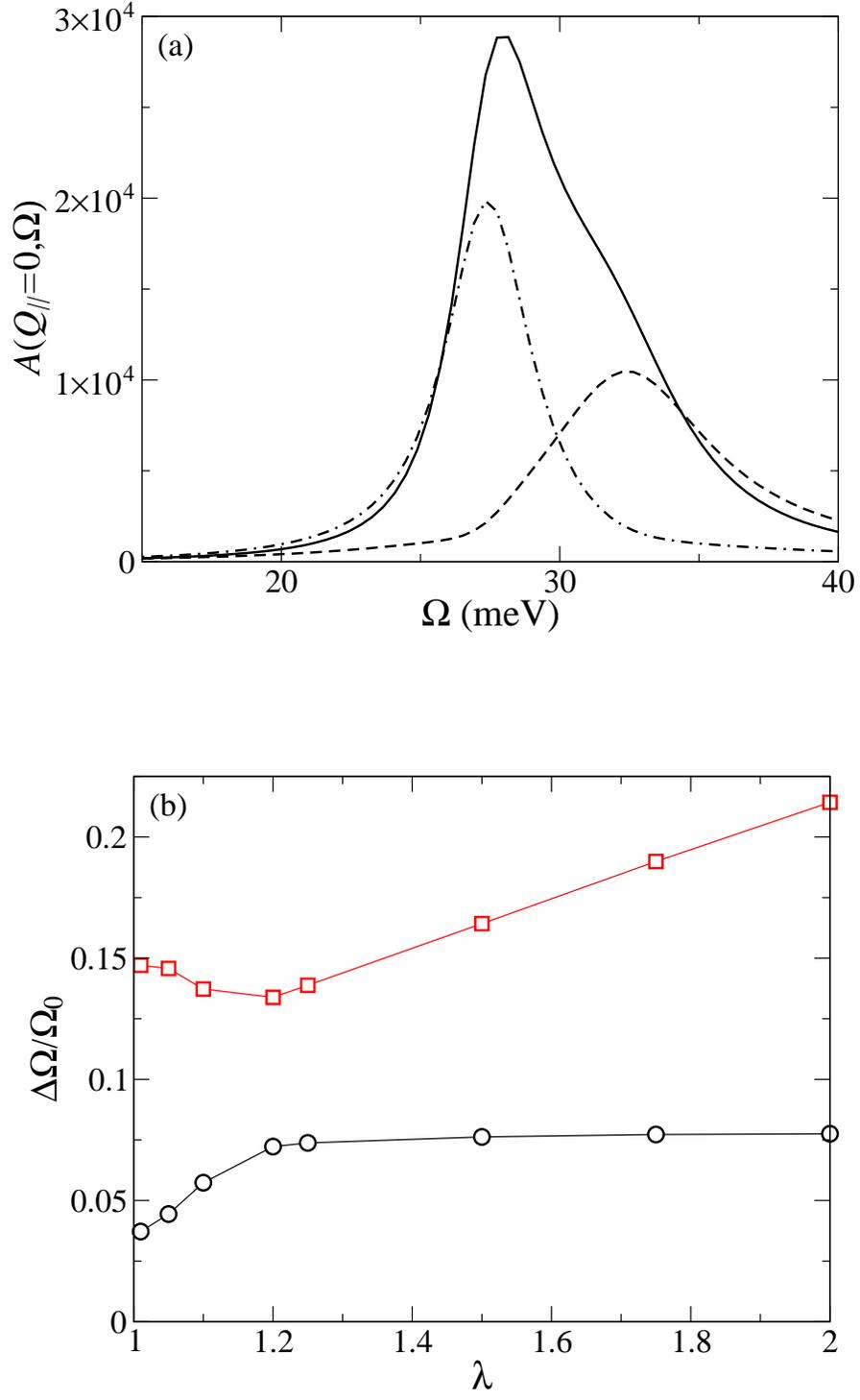

\begin{center}
\epsfig{file=fig8a.eps,scale=.5}\vspace*{.75in}\\
\epsfig{file=fig8b.eps,scale=.5}
\end{center}
\caption{This figure gives information on the
spectrum of a Co$_2$Cu$_{10}$Co$_2$/Cu(100)
structure, where the frequency of the outer film
is ‡0 and that of the inner film is $\Omega_0$.
In (a), for the case where $\lambda$ is 1.2, we
show with the solid line the spectral density
function $A(\Omega )$ defined in the text.  The
dashed line is the spectral function $A_1(\Omega
)$ and the dot dash line is $A_2(\Omega )$. In
(b) we give the linewidth of the acoustical and
optical modes of the trilayer as a function of
the parameter $\lambda$.}
\end{figure}

\begin{thebibliography}{99}
\bibitem{18}
Quite a number of authors use an adiabatic
approach to calculate effective exchange
couplings between nearby spins in ultrathin
ferromagnets, then generate spin wave spectra
utilizing a Heisenberg Hamiltonian appropriate to
localized spins. In this picture, for a given
wave vector in the surface Brillouin zone, one
realizes $N$ spin wave modes, each with infinite
lifetime, with $N$ the number of layers in the
film. The calculations reported in ref. [15] and
ref. [16] show that the adiabatic approximation
breaks down qualitatively for ultrathin
ferromagnetic films, and the picture of spin
excitations just described is qualitatively
incorrect. Instead, through most of the surface
Brillouin zone, save for the near vicinity of the
point, our theory predicted that one has a
single, broad feature in the frequency spectrum
of spin fluctuations, which displays dispersion
with wave vector similar to that associated with
a single spin wave branch. This picture is
confirmed by the data reported in Ref. [17], and
in Ref. [16] we demonstrate our approach provides
a fully quantitative account of the data. An
explicit comparison between adiabatic or
Heisenberg like calculations, and our full
dynamical calculations see Fig. 4 and Fig. 5 of
Ref. [16].
\end{thebibliography}
\end{document}